# Heated gas bubbles enrich, crystallize, dry, phosphorylate, and encapsulate prebiotic molecules

Matthias Morasch*, Jonathan Liu*, Christina F. Dirscherl*, Alan Ianeselli*, Alexandra Kühnlein*, Kristian Le Vay§, Philipp Schwintek*, Saidul Islam+, Mérina K. Corpinot+, Bettina Scheu#, Donald B. Dingwell#, Petra Schwille§, Hannes Mutschler§, Matthew W. Powner+, Christof B. Mast*, and Dieter Braun*

\* *Physics Department, NanoSystems Initiative Munich and Center for Nanoscience Ludwig-Maximilians-Universität München, Amalienstrasse 54, 80799 München, Germany*
\# *Earth and Environmental Sciences, Ludwig-Maximilians Universität München, Theresienstr. 41, 80333 München, Germany*
§ *Max Planck Institute of Biochemistry, Am Klopferspitz 18, 82152 Martinsried, Germany*
\+ *Department of Chemistry, University College London, 20 Gordon Street, London, WC1H 0AJ, UK*

**Abstract.** Non-equilibrium conditions must have been crucial for the assembly of the first informational polymers of early life—by supporting their formation and continuous enrichment in a long-lasting environment. Here we explored how gas bubbles in water subjected to a thermal gradient, a likely scenario within crustal mafic rocks on the early Earth, drive a complex, continuous enrichment of prebiotic molecules. RNA precursors, monomers, active ribozymes, oligonucleotides, and lipids are shown to (1) cycle between dry and wet states, enabling the central step of RNA phosphorylation, (2) accumulate at the gas-water interface to drastically increase ribozymatic activity, (3) condense into hydrogels, (4) form pure crystals, and (5) encapsulate into protecting vesicle aggregates that subsequently undergo fission. These effects occurred within less than 30 minutes. The findings unite physical conditions in one location which were crucial for the chemical emergence of biopolymers. They suggest that heated microbubbles could have hosted the first cycles of molecular evolution.

Life is a non-equilibrium system. By evolution, modern life has created a complex protein machinery to maintain the nonequilibrium of crowded molecules inside dividing vesicles. Based on entropy arguments, equilibrium conditions were unlikely to trigger the evolutionary processes during the origin of life[1]. External non-equilibria had to be provided for the accumulation, encapsulation, and replication of the first informational molecules. They can locally reduce entropy, give rise to patterns[2], and lean the system towards a continuous, dynamic self-organization[3]. Non-equilibrium dynamics can be found in many fluid systems, including gravity-driven instabilities in the atmosphere[4], the accumulation of particles in nonlinear flow[5,6], and shear-dependent platelet activation in blood[7]. Our experiments discuss whether gas-water interfaces in a thermal gradient could have provided such a non-equilibrium setting for the emergence of life on early Earth.

Non-equilibrium systems in the form of heat flows were a very common and simplistic setting, found ubiquitously on the planet[8]. Hydrothermal activity is considered abundant on early Earth and intimately linked to volcanic activity[9]. Water is thereby circulating through the pore space of the volcanic rocks, which is formed by magmatic vesiculation (primary origin) and fractures (secondary origin). These systems have been studied as non-equilibrium driving forces for biological molecules in a variety of processes[10-17].

Gases originating from degassing of deeper magma bodies percolate through these water-filled pore networks. At shallow depths bubbles are formed by gases dissolved in water and formation of vapor where sufficient heat is supplied by the hydrothermal system. The bubbles create gas-water interfaces, which previously have been discussed in connection with atmospheric bubble-aerosol-droplet cycles[18], the adsorption of lipid monolayers and DNA to the interface[19,20], or the formation of peptide bonds[21].

In the absence of a temperature gradient, the evaporation of a drop of water on a surface exhibits the so-called "coffee-ring effect"[22]. Upon the evaporation, molecules in the drop are accumulated at its rim by a capillary flow. After complete evaporation, a ring of concentrated material is deposited. In the inverted setting studied here, a gas bubble is immersed in water (Fig. 1) and a temperature gradient drives this process continuously.

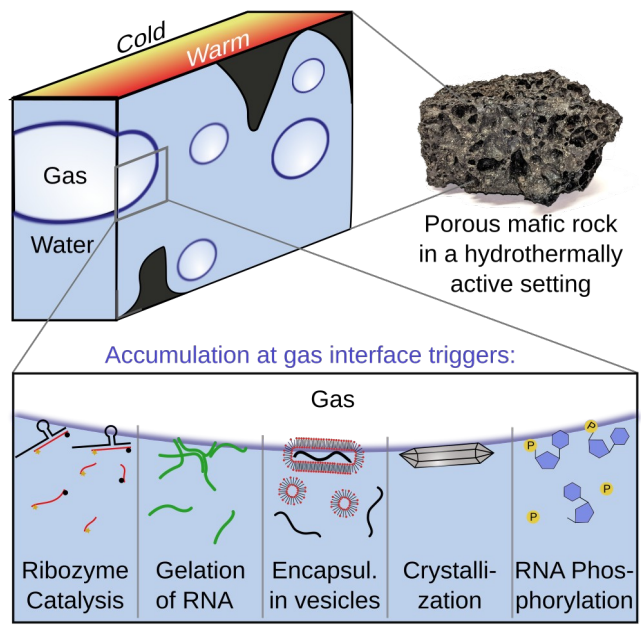

*Figure 1: DNA accumulation at gas bubbles in a thermal gradient. Volcanic rocks in shallow hydrothermal settings are subjected to water cycles in the pore space of both primary (magmatic vesicles) and secondary (fractures) origin. Gases originating from magma degassing at depth percolate through the water. The heat supplied by the hydrothermal system causes vaporization. At the gas microbubbles, molecules are accumulated by the continuous capillary flow on the warmer side of the gas-water interface. As shown experimentally, this environment can enhance the catalytic activity of ribozymes, trigger the formation of a hydrogel from self-complementary RNA, encapsulate oligonucleotides such as aptamers in vesicle aggregates, trigger their subsequent fission, drive the crystallization of ribose aminooxazoline (RAO)—a prebiotic RNA precursor—and initiate the phosphorylation of RNA nucleosides.*

**Results—Accumulation at the gas-water interface.** Experimentally, bubbles were created by filling a 240 µm thick, corrugated microfluidic chamber with solution (Supplementary Fig. 1). As the solution could not fill all cavities, pinned gas bubbles were created. At higher temperatures, bubbles were found to also form spontaneously anywhere in the system from outgassing. These bubbles were often not restricted by their surrounding geometry and moved along the heated surface.

The accumulation of molecules at these heated gas-water interfaces is caused by a continuous evaporation-recondensation cycle of water. The interfaces were held in a constant non-equilibrium state, leading to a steady state coffee-ring effect that does not end in a fully dry state. Here we observe six physico-chemical processes, which have all individually been suggested to be relevant for the emergence of prebiotic evolution and are co-located in a single non-equilibrium system: (i) enhanced catalytic activity of ribozymes, amplified by the accumulation of oligonucleotides and ions, (ii) condensation of self-complementary RNA 36mers into millimeter-sized hydrogels, (iii) vesicle aggregation at the bubble interface along with encapsulation of oligonucleotides in aqueous phases with up to 18-fold enhanced concentration, (iv) fission of the vesicle structures in the adjacent micro-convection, (v) formation of euhedral 300 µm crystals from the RNA precursor sugar ribose aminooxazoline (RAO) around bubbles, which also act as seeds for new bubbles, and (vi) dry-wet cycles enhancing for example the phosphorylation of nucleosides, created by fluctuating and moving interfaces (Fig. 1, Supplementary Movies M1-M6). All six mechanisms were established within 30 minutes and, importantly, operated in continuous contact with bulk water.

Here, length- and temperature-dependent accumulations were measured at low-salt conditions (0.1-fold PBS buffer: 13.7 mM NaCl, 0.27 mM KCl, 1 mM phosphate buffer). However, DNA and RNA gelation measurements were performed equally well under physiological conditions (1-fold PBS buffer: 137 mM NaCl, 2.7 mM KCl, 10 mM phosphate buffer).

In order to observe the dynamics at the interface, bubbles were created as described above using fluorescently labeled molecules in solution (Fig. 2a(i)). The front- and back-side of the chamber were heated and cooled, respectively, to generate a temperature gradient. The system was monitored through the warm side using a fluorescence microscope. The optical axis runs along the temperature gradient (Supplementary Fig. 1,2) and the quantitative fluorescence was captured with a CCD camera.

Initially, the chamber was filled with a solution of 200 nM FAM-labeled 132 base single-stranded DNA (ssDNA) oligomer in 0.1-fold PBS buffer. When no temperature gradient was applied to the system ($T_{warm} = T_{cold} = 10$ °C), we observed no accumulation of DNA near the gas-water interface. The fluorescence signal exhibited a constant small peak at the observed interface, possibly due to a slight adsorption of the DNA to the gas-water interface (Fig. 2a(i), 0 s). Heating one side of the chamber ($T_{warm} = 30$ °C, $T_{cold} = 10$ °C) resulted in the rapid accumulation of DNA in a small area on the warm side at the contact line (Fig. 2a(ii), dashed red box, Supplementary Movie M1). The chamber-averaged fluorescence at the contact line increased within six minutes approximately 12-fold compared to the bulk fluorescence (Fig. 2b). No accumulation was observed on the cold side.

We calculated the local concentration at the contact line from the ratio of the meniscus- and bulk-fluorescence and the geometry of the curved gas-water interface. Since the fluorescence was averaged over the chamber by the microscope objectives, a 60-fold higher concentration in addition to the higher fluorescence is inferred due to the thinner size of the accumulation region (approximately 4 µm) as compared to the 240 µm wide chamber. Therefore, from the observed 12-fold increase in fluorescence, we estimated a concentration increase by a factor of 700, corresponding to 140 µM DNA concentration in the meniscus when starting from a 200 nM bulk solution (Fig. 2c). Simulations suggested that without thermophoresis this accumulation would be only slightly higher, showing that it does not play a significant role in the accumulation process (Fig. 2c, red interrupted line).

Further analysis from experiment and theory showed that the accumulation was caused by the focused evaporation of water at the tip of the meniscus[23] (Fig. 2d, orange). A continuous flow of water into the meniscus dragged the molecules with it, which, since they could not evaporate, could only escape by diffusion against the one-way capillary flow. The flow was visualized by filling the chamber with a suspension of 200 nm diameter FAM-labeled polystyrene beads. By particle tracking, we measured the velocity profile in the meniscus (Fig. 2d,e, Supplementary Movie M1). Beads moved towards the accumulation region near the hot side of the chamber at the contact line.

We attribute this to capillary flow, which superseded the comparatively weak bulk buoyant convection (Fig. 2e). Temperature gradients have also been demonstrated to create Marangoni flows[23], in which water is drawn from the warm to the cold side of an interface due to a surface tension gradient. This was difficult to observe because the main temperature gradient was along the viewing axis. However, we observed strong lateral flows at the interface when accumulating vesicles (see below). We attribute these to lateral Marangoni flows and therefore assume also a combination of Marangoni flows and convection along the interface (Fig. 2d, green). Without this flow our simulation predicts a larger accumulation (Fig. 2c, blue interrupted line).

The water that evaporated near the warm chamber wall was found to condense on the cold wall, forming small water droplets. As a result, the gas bubble had less space, expanded, and moved the gas-water interface. Once the condensed "rain" droplets had grown, merged, and re-entered into the bulk solution, driven by surface tension, the interface moved back to its initial position (Supplementary Movie M1). These fluctuations of the contact line triggered a drying and recondensation of the molecules at the location where the capillary flow initially accumulated them.

A fluid dynamics model described the main features of the accumulation. We considered a 2D section perpendicular to the interface. Since the width of the channel was smaller than the capillary length for water and gas[24], the interface geometry was approximated by an arc of a circle. The contact line of the interface was pinned in the simulation[25,26], motivated by the observation that the accumulation kinetics was generally faster than the fluctuating movement of the interface.

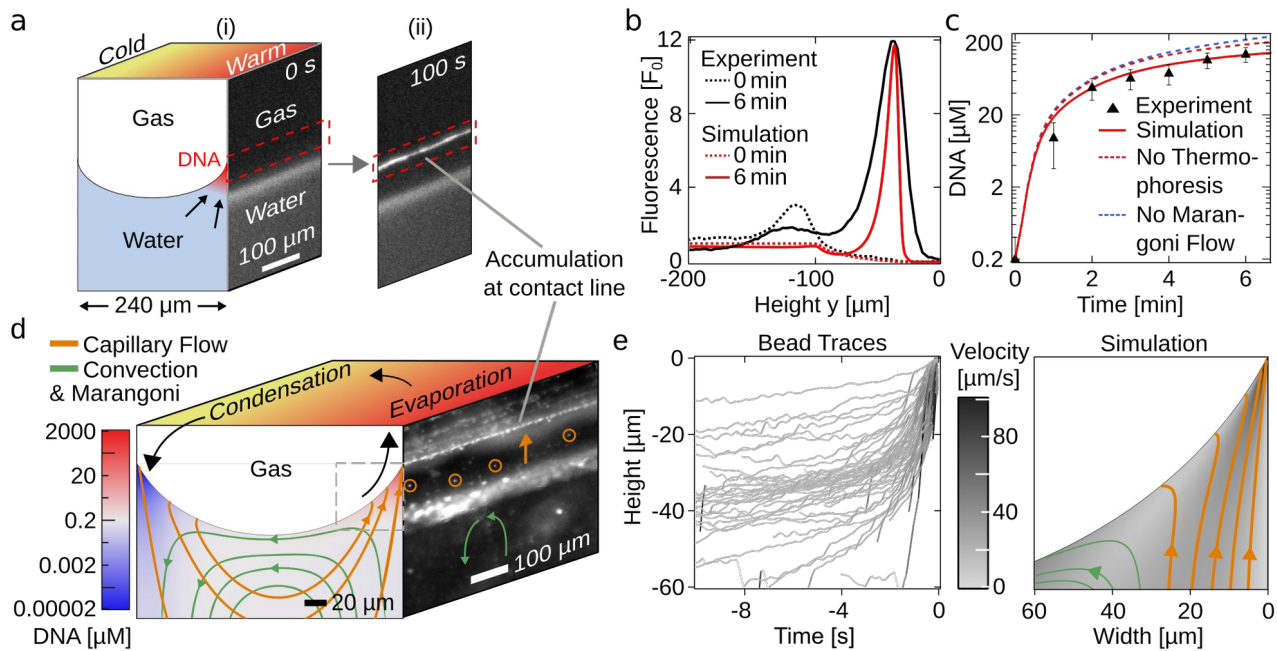

*Figure 2. DNA accumulation by capillary flow. (a) (i) gas-water interface imaged by fluorescence microscopy with the contact line area marked in red. The initial higher fluorescence intensity at the contact line originated from adsorbed DNA at the gas-water interface. (ii) After 100 seconds, the 132mer DNA accumulated in water near the contact line at the warm temperature side of the chamber ($T_{warm}$ = 30 °C, $T_{cold}$ = 10 °C, Movie M1). (b) The fluorescence profile reached a 12-fold increase as compared to the bulk fluorescence within six minutes. (c) Due to the confined meniscus geometry, the accumulated concentration was significantly higher. The 12-fold increase in fluorescence corresponded to a 700-fold increase in DNA concentration, consistent with simulation results (red solid line). Without thermophoresis (red interrupted line) or Marangoni flows (blue interrupted line), simulations predict a slightly higher accumulation. Error bars were estimated from fluorescence analysis averaged over an approx. 200 µm large interface width at three positions along the contact line from the shown example measurement. (d) Fluid flow near the contact line measured by single particle tracking (Supplementary Movie M1). The capillary flow (orange) pulled the beads upwards toward the contact line as water mainly evaporated at the tip of the meniscus. Marangoni- and convection flows (green) shuttled the beads between hot and cold side. Superposed is the logarithmic concentration profile obtained from the simulation for the accumulation in (a). (e) Single-particle tracking of the capillary flow. Peak flow velocities reached 50 µm/s in the last second of the water flow before its evaporation (Supplementary Fig. 3). Here $T_{cold}$ = 10 °C, $T_{warm}$ = 40 °C.*

The model superposed four water flows that provide the boundary condition for the accumulation of DNA. They were (i) the capillary flow at the meniscus, (ii) the diffusion of water vapor between the interface and the gas bubble, (iii) the convection of water, and (iv) the Marangoni flow along the interface. The relative strength of the Marangoni flow–a free parameter due to the unknown presence of surface-active molecules[27]–was adjusted to fit the velocities measured in the experiment.

The interplay of all four flows led to the accumulation of molecules at the meniscus. As only water evaporated on the warm side, dissolved molecules were continuously dragged towards the contact line, where their concentration depended on their back-diffusion and the speed of the capillary flow. Convection and Marangoni flow provided a constant cycling of water and new material towards the accumulation region.

The DNA accumulation was experimentally measured for various temperature differences and DNA lengths (Supplementary Fig. 4). It rises with increasing temperature difference ($\Delta T$), reaching a 4000-fold increase for $\Delta T = 40$ °C. We also found that smaller (15mer) DNA accumulated three times less effective than larger (132mer) DNA, which was attributed to their higher diffusion coefficient. The model predicted a multi-fold accumulation of mono- and divalent ions (Supplementary Fig. 4), resulting in a higher salt concentration at the meniscus.

**Enhanced ribozyme catalysis at the interface.** The accumulation of larger biomolecules as well as ions at the interface makes it a powerful mechanism to enhance the activity of functional nucleic acids. To test this, we monitored the activity of the Hammerhead ribozyme[28,29], which cleaves a 12mer RNA substrate strand. The substrate and magnesium concentration determined its activity[30], and both were accumulated in a 30 °C gradient ($T_{warm}$ = 10 °C, $T_{cold}$ = 40 °C) at low bulk concentrations (0.1 µM Hammerhead, 0.5 µM Substrate, 0.4 mM $MgCl_2$, Fig. 3a). As a control, chambers without gas interfaces were studied. A FAM dye and a black hole quencher were attached on opposite sides of the substrate, inhibiting the fluorescence of the FAM dye. Upon cleavage, the dye was not quenched anymore and could be detected by fluorescence microscopy. Fig. 3a,b and Movie M2 show the average fluorescence of the chamber over time. The substrate strands were cleaved predominantly at the interface, as seen by the rise of fluorescence there. From here, the cleaved strands were frequently ejected into the bulk solution. After 25 minutes, samples were extracted from the bulk fluid and analyzed by polyacrylamide gel electrophoresis (PAGE, Fig. 3c). We detected up to 50 % concentration of the cleaved substrate (bottom band, Supplementary Fig. 5) in a chamber with bubbles.

In the same temperature gradient, but without interfaces, only 3.8 % of the substrate was cleaved. This decreased even further when the bulk temperature was set homogeneously to 40 °C, where only very little activity could be observed. When using an inactive mutant version of the ribozyme, the fluorescence did not noticeably increase (Fig. 3a, top right, Fig. 3b dotted line), demonstrating that it did not originate from an enhanced hydrolysis at the interface. This shows that the accumulation mechanism can enhance the catalytic activity of ribozymes, while maintaining low salt conditions in the bulk solution. It increases the turnover, i.e. the number of strands cleaved per ribozyme, by the accumulation of substrates and ions.

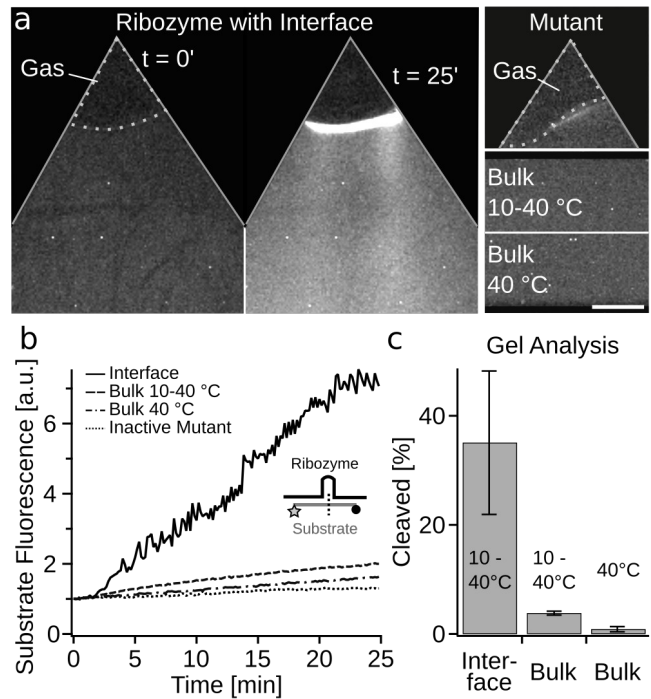

*Figure 3: Ribozyme catalysis triggered by interface.* Fluorescence microscopy and PAGE analysis of Hammerhead ribozyme activity. *(a)* Hammerhead in a chamber with gas-water interface after 0 and 25 minutes (left and right, respectively, $T_{warm}$ = 40 °C, $T_{cold}$ = 10 °C, 0.1 µM Ribozyme, 0.5 µM Substrate, 0.4 mM $MgCl_2$). The fluorescence increased strongly at the interface over time (Supplementary Movie M2). Without interface (bottom right), the Hammerhead shows little activity, both in the same temperature difference and at a homogeneous 40 °C; an inactive mutant (top right) showed no significant increase in fluorescence. Scale bar: 500 µm. *(b)* Overall fluorescence in the bulk fluid from (a) over time. The illustration shows the ribozyme (black) and substrate (grey) with the cleavage site (dotted line), dye (star), and quencher (black circle). *(c)* PAGE analysis results with and without interface after 25 minutes (gels in Supplementary Fig. 5). We found a significant increase in cleaved product with the interface (35 % ± 13 %, five measurements) compared to bulk samples in the same temperature gradient (3.79 % ± 0.35 %, three measurements) or at 40 °C (0.86 ± 0.49, three measurements). Errors are standard deviations from normalized gel band intensities, repeats are given above.

***RNA/DNA gelation.*** The length-selectivity observed in nucleic acid accumulation could increase the concentration of self-complementary oligonucleotides to the point that hydrogel-forming concentrations could be reached (Fig. 4a). Self-complementary strands were found to form a macroscopic, millimeter-sized hydrogel, a process previously shown for DNA in a thermophoretic accumulation chamber[15]. We observed the formation of hydrogels for self-complementary DNA and RNA and both for GC-only (Fig. 4, Supplementary Movie M3) and AT-only sequences (Supplementary Fig. 6). All oligonucleotides were end-labeled and HPLC purified to minimize the presence of free dye in the experiments.

Starting from uniformly distributed DNA in bulk solution (Fig. 4a(i), initial concentration 10 µM; 1-fold PBS buffer), the DNA was quickly accumulated at the interface once the temperature gradient was established. Within eight minutes, a hydrogel had formed, after which it quickly detached from the interface and entrained in the convection flow (Fig. 4a(ii)). The hydrogel nature of the DNA was checked by increasing the temperature from 30 °C, below the melting point of the self-complementary sequences, to 70 °C, where the hydrogel dissolved into the convection flow (Fig. 4a(iii)).

The formation of hydrogels from self-complementary RNA was also observed. Here we co-accumulated two different sequences in a 20 °C temperature difference (Fig. 4b). The red fluorescence channel monitored a 36 mer non-complementary ssRNA strand and the green fluorescence channel a self-complementary GC-only 36 mer ssRNA strand with three self-complementary binding sites. Based on simulations (Nupack, www.nupack.org), the hydrogel-forming strands bind to each other and form a network of polymers (Fig. 4b, right). Over the course of the experiment, both the non-complementary and self-complementary strands accumulated at the same interface. But, after 21 minutes, we only observed a hydrogel for the self-complementary green RNA. No hydrogel could be observed for the non-complementary red RNA, which also accumulated near the surface but was not forming large-scale structures (Supplementary Movie M3). Replacing the GC-only strand with a 60mer AU-only RNA with similar self-complementarity gave the same results (Supplementary Fig. 6). This demonstrates that the gel-formation and separation of the strands is not dominated by G-quadruplex formation.

The red and green strands separated macroscopically based only on their sequence. The self-complementary strands remained at a local high concentration in the hydrogel, which offered reduced hydrolysis rates due to its predominantly double-stranded nature. This sequence selective gelation was similarly found also for DNA (Supplementary Fig. 6).

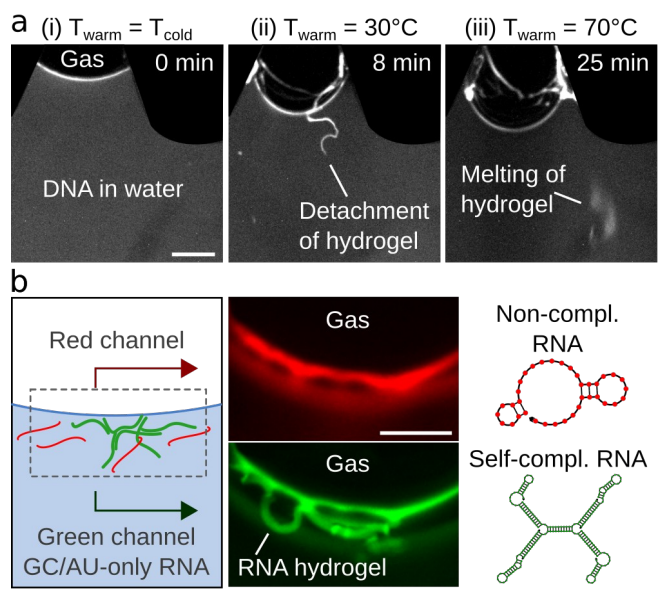

*Figure 4. Sequence selective gelation of RNA and DNA. Fluorescence microscopy of DNA and RNA revealed the formation of hydrogels at the interface. (a) Gelation of DNA. (i) Before applying the temperature gradient, DNA accumulated at the interface only due to slight surface adsorption to the gas-water interface. (ii) After applying the temperature gradient, a DNA hydrogel formed and detached from the interface (Supplementary Movie M3). (iii) At $T_{warm}$ = 70 °C, the hydrogel melted and redistributed the DNA back into the bulk fluid. We estimated the DNA concentration in the hydrogel to be 100 µM. Scale bar 500 µm. (b) Gelation of RNA. In a single experiment, non-complementary 36 mer ssRNA (red) was accumulated with self-complementary GC-only 36 mer ssRNA (green). Both strands accumulated at the interface ($T_{warm}$ =30 °C, $T_{cold}$ = 10 °C), however only the self-complementary RNA formed an elongated, fibrous hydrogel. The same behavior was found for DNA and AU-only RNA (Supplementary Fig. 6). Scale bar 125 µm.*

***DNA Encapsulation in vesicle aggregates.*** A key requirement for the emergence of cellular life is the encapsulation of molecules at increased concentration relative to their more dilute external environment. Fatty acids are potential candidates that could separate nucleic acids in vesicles, possibly incorporating phospholipids into their membranes over time[31,32]. For the encapsulation, three autonomous processes need to occur: (i) The accumulation of oligonucleotides to meaningful concentrations, (ii) the accumulation of vesicles to trigger their aggregation or fusion, and (iii) the combination of both in one location to encapsulate oligonucleotides into vesicular structures. Here we show that heated gas-water interfaces could fulfill these requirements. The accumulated vesicles not necessarily form larger, round vesicles, but aggregates. However, we demonstrate below that these aggregates enclosed DNA and RNA in an aqueous phase, allowing their binding and folding. Shear flows close to the interface as well as in the convection flow were shown to divide these aggregates to form smaller and more round structures.

To introduce lipids in a homogeneous manner, the chamber was filled with small 100 nm-sized vesicles prepared from a 10 mM oleic acid solution (0.2 M Na-Bicine, 1 mM EDTA, pH 8.5) and 2 µM DNA. These initially small vesicles appeared as a continuous background (Fig. 5a,b, Supplementary Movie M4). After turning on the temperature gradient, we observed the accumulation of these vesicles together with DNA within ten minutes at the interface ($T_{warm}$ = 70 °C, $T_{cold}$ = 10 °C). Here, the vesicles aggregated together and formed larger clusters. It should be noted that this aggregation and cluster-formation was strongly increased if approximately 0.1 % 1,2-dioleoyl-sn-glycero-3-phosphoethanolamine (DOPE) was added to the lipids, present here also to label the lipids. Interestingly, the co-accumulated DNA was encapsulated into these vesicle aggregates, in which we found an up to 18-fold increase in oligonucleotide concentration compared to the bulk solution. These aggregates were shuttled into the convection and frequently formed thread-like structures. Close to the interface, we observed strong flows (Supplementary Movie M4) which we attribute to Marangoni flows. These could originate from lateral temperature gradients across the chamber due to inhomogeneous heating or differences in thermal conductivity of the chamber material and water. Aggregates were observed to divide and split into smaller compartments in these flows and the convection (Supplementary Fig. 4).

Vesicles formed from the phospholipid DOPC (3.6 mM DOPC, 0.2 M Na-Bicine, 1 mM EDTA, pH 8.5) showed a similar behavior, however they aggregated stronger at the interface and externally applied flow across the microfluidics was sometimes necessary to remove them from the interface. The accumulated 100 nm-sized vesicles formed larger and more spherical structures compared to the oleic acid aggregates, and encapsulated DNA equally well. They also underwent fission due to shear stress in the convection flow (Fig. 5c, Supplementary Movie M4).

The vesicle aggregates (oleic acid or DOPC) concentrated around smaller gas-bubbles (150 µm in diameter), inducing a clustering of DNA (Supplementary Fig. 7). This was not observed in the absence of lipids. Lipid vesicles therefore significantly enriched the local DNA concentration.

The co-location of DNA and lipids raises the question whether the oligonucleotides are in an aqueous phase that allows e.g. the folding of RNA or binding of DNA, and if these compartments are protected from their surrounding. To test the former, we accumulated the RNA aptamer "Broccoli"[33] (Fig. 5d), which folds around the fluorophore DFHBI-1T and increases its fluorescence, with DOPC vesicles (1 µM Aptamer, 10 µM DFHBI-1T, 3.6 mM DOPC, 50 mM HEPES, pH 7.6, 100 mM KCl, 1 mM $MgCl_2$). Here, $T_{warm}$ = 40 °C and $T_{cold}$ = 10 °C to avoid RNA and fluorophore degradation. Again, the RNA, fluorophore, and vesicles accumulated at the interface, leading to the formation of aggregates that were visible both in the lipid as well as the fluorophore color channel. Replacing the Broccoli aptamer with a nonbinding RNA strand led to a more than 100-fold reduced fluorescence of the fluorophore. This shows that the aptamer was folded inside the aggregates.

To demonstrate that the accumulated material was protected inside the aggregates, a 72mer double-stranded DNA (dsDNA) was accumulated with DOPC vesicles (Fig. 5e, 7.1µM DNA, 3.6 mM DOPC, 0.2 M Na-glycineamide pH 8.5, 6 mM $MgCl_2$, 1 mM $CaCl_2$). Each pair of dsDNA strands thereby contained a FAM dye on one strand and a ROX dye opposite to it on the other strand. These dyes form a Förster resonance energy transfer (FRET) pair, in which the excited FAM dye can transfer energy to the ROX dye that then fluoresces. As this energy transfer works only in close proximity of the dyes, we could use it to measure the amount of dsDNA inside the vesicle aggregates. The higher the signal—between 0 and 1—the more dsDNA was present. After the formation of aggregates at the interface, most of the solution was extracted and DNase I added to it (0.5 units to approx. 10 µl). The solution was then filled back into the chamber and the FRET signal observed at 37 °C. The DNase digested the DNA strands outside the aggregates, where the signal quickly dropped (Fig. 5e, Supplementary Movie M4, Supplementary Fig. 7). The FRET signal of the aggregates reduced only slightly, likely from the digestion of solution above/below or DNA sticking to their surface. We could thereby show that the aggregates protect the molecules inside them.

The FRET signal was also used to observe the melting of dsDNA inside the chamber (Fig. 5f). After accumulation (7.1 µM DNA, 3.6 mM DOPC, 0.2 M Na-Glycineamide pH 8.5, 11 mM NaCl, 0.22 mM KCl, 0.8 mM phosphate buffer), the chamber was heated to 95 °C, during which the FRET signal reduced to 0 in the bulk and approximately 0.3 in the aggregates. The remaining signal could stem from aggregated DNA or DNA that is enclosed and stabilized by lipids in a way that did not allow the strands to fully unbind. After cooling, the FRET signal returned to 1 for both aggregates and bulk solution. The combination of aptamer and FRET analysis demonstrates that oligonucleotides are encapsulated inside the aggregates in an aqueous phase that allows them to melt and fold.

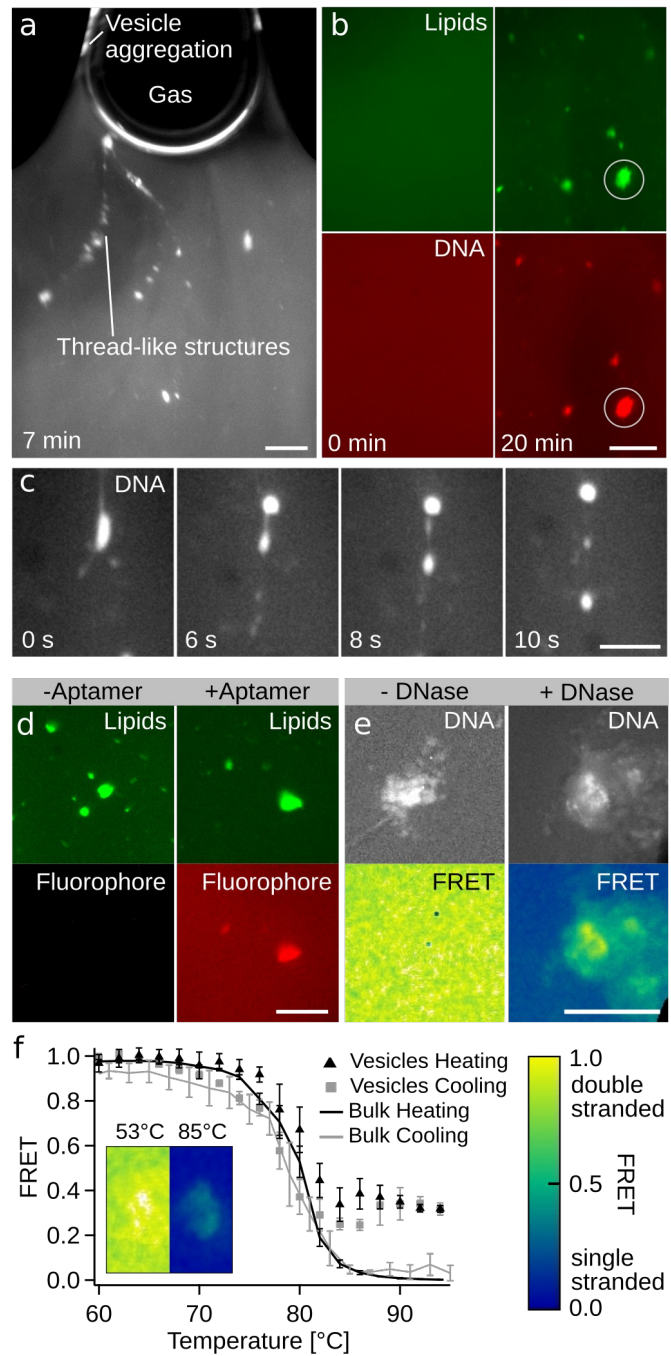

*Figure 5. DNA and RNA encapsulation and protection in vesicles formed at the interface. Fluorescence microscopy images of DNA/RNA and lipid color channels. (a) Lipid channel: 100 nm sized oleic acid vesicles were accumulated at the gas-water interface. They formed aggregates of vesicles that were ejected into the bulk and formed thread-like structures (Supplementary Movie M4). (b) Comparison of lipid and DNA color channels (green: lipid, red: DNA) showed that lipids and DNA co-localize, meaning that the 100 nm-sized vesicles formed larger clusters and encapsulated DNA at up to 18-fold enhanced concentration compared to the bulk. (c) DNA channel:*

*DOPC vesicles were accumulated under the same conditions and also encapsulated DNA at enhanced concentrations. Here, a fission of the DOPC vesicles containing DNA was observed in the shear flow. **(d)** Encapsulation of the fluorophore DFHBI-1T with a non-binding RNA (left) and the Broccoli aptamer (right) that increases its fluorescence. It shows that RNA folded in an active conformation inside the vesicles. **(e)** FRET analysis of dsDNA in aggregates before (left) and after (right) addition of DNase I. DNA inside the aggregates was protected from the DNase (high FRET), while it was digested in the bulk solution (low FRET, Supplementary Movie M4). **(f)** FRET melting curve of dsDNA inside and outside the aggregates. Starting from 1 (all DNA double stranded) at 60 °C to 0 (all DNA single stranded) at 95 °C outside and 0.3 inside the aggregates. Error bars are standard deviations measured for a larger area of normalized signal of the shown example measurement. Scale bars: a,b: 200 µm, c: 100 µm, d,e: 250 µm.*

Above experiments assumed the presence of uniformly 100 nm-sized vesicles at the beginning. We also explored the behavior of the system when it initially contained a range of vesicle sizes with up to approximately 30 µm in diameter (Supplementary Fig. 8). They exhibited a similar, but often slower accumulation behavior. Within 20 minutes, the system again started to form vesicle clusters that contained enhanced DNA concentrations. The formation of oleic acid aggregates was also observed in the absence of DNA, indicating that they were not the result of DNA/lipid interaction (Supplementary Fig. 8).

*Crystallization at gas bubbles.* The building blocks for the synthesis of single nucleotides, such as the prebiotic RNA precursor ribose aminooxazoline (RAO), accumulated near gas bubbles to concentrations that triggered its crystallization. For RAO, a crystallization is of fundamental interest since it can be both diastereoisomerically purified by selective sequential crystallization of its precursors and enantiomerically enriched by conglomerate crystallization where the two enantiomers (D- and L-) of RAO crystallize into discrete independent domains[34,35]. To trigger a controlled crystal growth, RAO would need to be accumulated slowly around a growing bubble. In previous experiments, bubbles were artificially created. Here we used elevated temperatures (70 °C warm side, 10 °C cold side) to trigger the spontaneous formation of a bubble. Subsequently, RAO accumulated and crystallized around it (Fig. 6a,b). Therefore, a 40 mM solution of D-RAO, 2.8-fold below the saturation concentration at $T_{warm}$ = 70 °C (approximately 110 mM), was filled in a chamber without a corrugated geometry and did not create a gas-water interface. For the crystallization to occur at the warm side of the chamber, RAO would have needed to accumulate several fold in order to overcome the nucleation energy barrier[36]. We monitored the fluorescence of 1 µM Cy5 added to the solution, which co-accumulated at the gas-water interface but was not incorporated into the crystals.

An initially small bubble formed on the warm side and accumulated RAO around it. Within 40 minutes, the bubble grew while continuously increasing the RAO concentration at its warm side. Here, the crystal shown in Fig. 6a was found at the location where the bubble had formed. No crystals were found on the cold side of the chamber. An X-ray crystal structure determination confirmed that crystals grown on the warm side were indeed D-RAO (Fig. 6c).

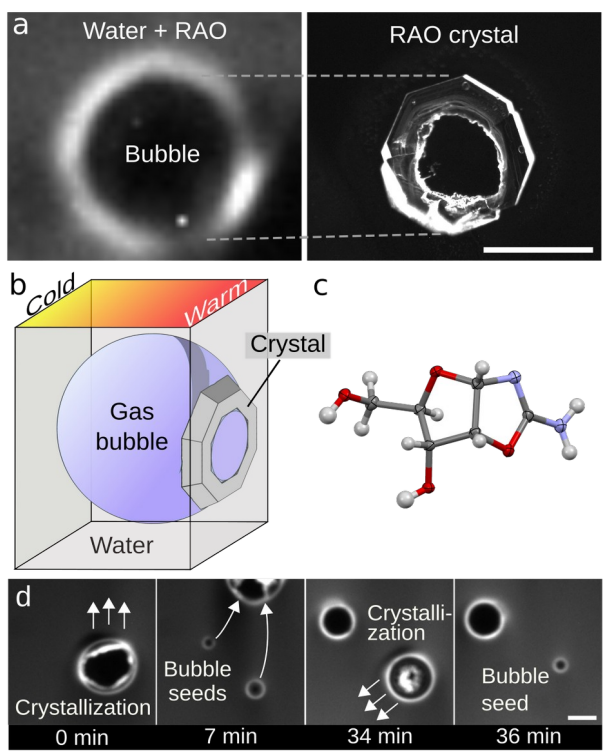

*Figure 6: Crystallization and bubble movement. (a) RAO accumulated from 40 mM bulk concentration around a growing bubble at the warm side of the chamber (70 °C). After 37 minutes, an RAO crystal formed below the bubble at the water-gas-solid interface as indicated. (b) The crystal shape suggests that it formed successively as the bubble grew, consistent with simulation results (Supplementary Fig. 9). (c) X-ray crystal structure analysis of a D-RAO crystal grown at a gas-water interface. (d) Moving bubbles leave behind crystals (7 min) that act as seeds for new bubbles (34 min, Supplementary Movie M4). Scale bars: 250 µm.*

Interestingly, the formation of RAO crystals led to the later re-formation of gas-bubbles near small cavities—a known and well-described process of heterogeneous nucleation in supersaturated solutions[36]. We observed this crystal-induced bubble formation in experiments at high $T_{warm}$ = 70 °C (Fig. 6d, Supplementary Movie M4). Once a gas bubble, around which RAO crystals had formed, moved away, it left behind accumulated material, possibly crystal cavities still filled with gas, which facilitated the growth of new daughter bubbles at the same location. When these daughter bubbles also increased in size sufficiently, they started to accumulate RAO, formed crystals, and eventually moved away. The remaining crystals again hosted the growth of new bubbles and began the bubble-induced crystallization cycle again.

***Dry-wet-cycles and phosphorylation by moving interfaces.*** If bubbles were not confined by their geometry, they tended to move upwards in the chamber (Fig. 7a) due to buoyancy forces. This movement led to a continuous cycling of dry-wet conditions (Fig. 7a), as accumulated material close to the interface entered the bubble, dried, and was rehydrated when the bubble moved away. At the same time, accumulated material at the trailing edge was dragged along if the bubble moved slow enough, keeping high molecule concentrations in the vicinity of the interface (Supplementary Movie M6).

At a hydrophilic surface material such as silicon dioxide (quartz) used on the cold side in the experiments before, the contact angles of the condensed droplets were small and they re-entered quickly into the bulk water, leading to only little movement of the gas-water interface. To trigger many wet-dry cycles, we placed a Teflon foil on the warm and cold sides of the chamber. This led to the formation of many droplets on the cold side (Fig. 7b, small round structures inside the dark gas-region, Supplementary Movie M6) that, if large enough, could also be in contact with the warm side. In this setting, we tested the phosphorylation of nucleosides, a reaction which usually requires dry conditions at elevated temperatures. The reaction is known to be most effective when the solution is dried at 100 °C, a scenario hard to reconcile with typical RNA-world conditions[37,38]. We observed a twenty times more efficient phosphorylation of cytidine nucleosides (Fig. 7c, $T_{warm}$ = 60 °C, $T_{cold}$ = 37 °C, 240 mM cytidine & ammonium dihydrogen phosphate, 2.4 M urea, similar to the phosphorylation used in [37]) compared to bulk water at average temperatures of 50 °C and 60 °C. The found enhanced in situ phosphorylation reaction would improve the recycling of hydrolyzed RNA.

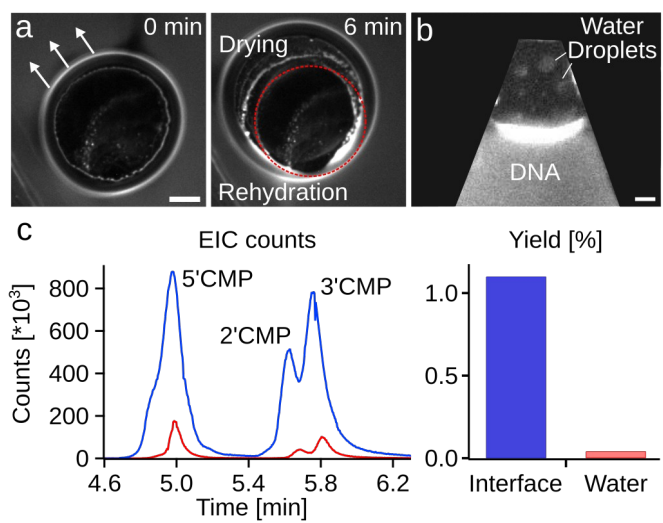

*Figure 7: Dry-wet cycles and phosphorylation of nucleosides. (a) The movement of bubbles in the chamber left behind dried material in the front that was rehydrated at the backside of the bubble, leading to dry-wet cycles (left: 0 min, right: 6 min). The dashed circle indicates the position of the accumulation region at t = 0 min. (b) Using a hydrophobic cold side in the chamber increased the speed of dry-wet cycling by enhancing the formation of circular water droplets on the cold side. As they "rained" into the bulk water, the interface moved periodically, triggering wet-dry cycles at the interface (Supplementary Movie M6). (c) Cytidine nucleosides were phosphorylated in the chamber shown in (b). Over 12h, the phosphorylation was on average twenty times more effective with the interfaces present (blue, $T_{warm}$ = 60 °C, $T_{cold}$ = 37 °C, CMP yield: 1.1 % ± 0.73 % from 12 repeats) compared to bulk water at 60 °C (red, CMP yield: 0.042 % ± 0.02 % from 5 repeats) or 50 °C (0.056 % ± 0.0015 % from 15 repeats). Left: Extracted-ion chromatogram counts (EIC) for CMP peaks after HPLC-MS analysis; right: Yields measured by HPLC and UV detection. Scale bars: 100 µm (a), 250 µm (b).*

*Discussion.* We found a general accumulation mechanism of molecules and small vesicles at gas bubbles subjected to heat flow in water. A temperature gradient across a gas-water interface created a continuous process of evaporation and condensation of water between the warm and cold sides. This moved molecules towards the bubble interface and increased their concentration by several orders of magnitude, depending on their diffusion coefficient. Since the contact line between gas and water was mobile, dry-wet cycles were created, in many conditions as often as twice per minute. The molecules studied here have been discussed as prebiotic candidates before the emergence of life[35,37,39]. Fluorescently labeled analogs were used to probe their concentration. We did not elaborate on the types of gases used in this study, since, due to their low concentration, the gases do not modify surface tension and water evaporation significantly[40].

Our simulation captured the basic characteristics of this accumulation and validated the experimental results. A 4000-fold DNA accumulation was reached both in experiment and simulation. In comparison with thermophoretic traps, with which the setting could be combined, the accumulation at the interface occurred significantly faster, on the timescale of minutes rather than hours[11].

We used different temperature gradients adapted to the different scenarios. The magnitude of accumulation dependent strongly on the applied temperature difference (Supplementary Figure 4), which was kept at high values in Figs. 3 and 4 (10 °C – 40 °C and 10 °C – 30 °C, respectively) to accumulate RNA strongly, but still under cold conditions to keep hydrolysis insignificant. Vesicles aggregated under similar conditions (Fig. 5d), but the faster convective flows from higher temperature difference (10 °C – 70 °C) shuttled aggregated vesicles more efficiently away from the interface for downstream analysis. The creation of crystals from RAO (Fig. 6) required the larger temperature gradients for an efficient accumulation of the comparably small molecules. The phosphorylation chemistry of monomers (Fig. 7) profited from an overall enhanced temperature.

The co-accumulation of small ions offer additional reactivity. Divalent salts such as $Mg^{2+}$ were predicted by our model to accumulate by a factor of 4 - 5 for temperature differences of 20 - 30 °C. In high salt environments, this could enhance the hydrolysis of for example accumulated RNA, however the larger temperature differences also increase the movement of the interface by the recondensation of water, decreasing the time accumulated molecules spend dried at high temperatures. On the other side, the enhanced salt concentrations would trigger ribozymatic activity at the interface while the molecules in the bulk are protected by low salt concentrations from hydrolysis.

The system provided in a single setting a network of widely different reactions, connected by the fast diffusive transport of these molecules through water between different microbubbles. Ligation[41] chemistry to drive replication could be made possible by the high local concentrations of substrates, whilst the drying process in close proximity would trigger the necessary phosphorylation[37,38] and activation chemistry[42] to drive the polymerization[43] of monomers. The dry-wet cycles were implemented continuously at the moving interface and molecules were retained at high concentrations at the interfaces after rehydration. In the adjacent water, molecules thermally cycle between warm and cold by laminar thermal convection.

At the interface, a strongly enhanced ribozymatic activity was demonstrated by the accumulation of the Hammerhead ribozyme. Its activity at the interface clearly dominates the turnover of substrate in the system, with up to 50 % of the product being cleaved in a chamber with the gas interface compared to 3.8 % in the bulk. This can be attributed to higher local concentrations of ribozyme, substrate, and $MgCl_2$ at the interface. Product strands were shuttled back into the bulk solution, where they are protected against the higher salt conditions. The mechanism thereby provides a way to enhance catalytic activity and increase the efficiency of RNA-catalyzed processes.

Complex sequence phenotypes of RNA with several self-complementary sites have shown a sequence selective formation of hydrogels at the interface. These hydrogels maintained a high local RNA concentration in water, an interesting setting to support efficient RNA catalysis[44,45] also since the high amount of hybridization in the hydrogels could protect oligonucleotides from hydrolysis[46] even in challenging salt concentrations.

In the presence of lipids such as oleic acid or DOPC, the accumulation at the heated gas-water interface led to a continuous encapsulation of oligonucleotides into vesicle aggregates. The local DNA concentration inside these structures increased by a factor of up to 18 as compared to the bulk solution. Folded RNA aptamers also accumulated inside the aggregates and dsDNA was shown to melt and re-anneal, demonstrating that aqueous phases readily exist inside the aggregates. As shown with DNase, the vesicles protected the encapsulated DNA from the bulk solution. The encapsulation of oligonucleotides into lipid membranes is considered to be one of the key elements for more complex life and it has been suggested that lipids facilitated the assembly and polymerization of monomers[47,48]. In the convection flow, the vesicles were subjected to temperature cycles and shear forces that lead to vesicle fission. DOPC also produced vesicular structures, showing that also modern phospholipids[39] could accumulate and encapsulate oligonucleotides in the shown conditions.

For the prebiotic synthesis of RNA, the crystallization at microbubbles would enable the purification of sugar-nucleobase precursors[34] and possibly also their chiral amplification by the enhanced growth of conglomerate RAO crystals[35]. Interestingly, the sites of crystal formation later triggered again the formation of gas bubbles, showing a self-selection for crystallization conditions. If the seed crystal was homochiral, the subsequent bubble formation could accumulate and promote the assembly of more homochiral molecules at the same location. Finally, we found that the prebiotically important dry chemistry of nucleoside phosphorylation was enhanced by the gas interface: Cytidine formed CMP twenty times more effective compared to aqueous conditions.

To conclude, the experiments showed multiple modes of condensation, enrichment, accumulation, and increased catalysis at heated gas microbubbles. This led to the physicochemical assembly and localization of prebiotic molecules – such as RNA precursors, lipids, and ribozymes. We argue that this accumulation of molecules at a gas-water interface was a robust feature of natural microfluidic systems in porous volcanic rocks in aqueous environments, a setting likely to be ubiquitous on early Earth[9]. The simultaneous occurrence of six synergistic mechanisms for the accumulation and processing of prebiotic molecules, all operated in close proximity, fulfills the requirements for early Life to connect a cascade of core reactions in the same non-equilibrium setting.

The setting presented here could therefore have largely helped in an informational polymer world, in which the first simple replicators were evolving. Following a synthesis of life's first building blocks, the accumulation at gas-water interfaces offers a mechanism to select polymers and enhance their catalytic activity. The shown continuous encapsulation dynamics of the accumulated molecules at the interface offer a pathway to emerge the cellular processes of life. Further experiments will test how this setting can host replication and selection towards early molecular evolution.

**Methods.** For the experiments in Fig. 2, a thermal chamber was sandwiched from a thin (>240 µm) layer of PETG plastic film deposited on a silicon wafer using an 3D printer (Ultimaker 2) in a funnel shape to facilitate a gas-water interface (Supplementary Fig. 1). A sapphire ($Al_2O_3$) block sealed the chamber. The chamber was filled through two 240 µm thick borosilicate capillaries (Vitrocom). The system was annealed at 150 °C and the sapphire pressed down, fixing the thickness to 240 µm. Polydimethylsiloxane (PDMS) was deposited around to seal the chamber. For most other experiments, chambers were built with a UV-curable resin (Photocentric 3D Daylight Resin, flexible, color: amber) through a master obtained by laser printing on a transparency film. Spacers between master and resin defined a chamber thickness to 150 µm or 250 µm. After illumination, sapphire windows were placed on top of the chambers and sealed with the resin. Microfluidic access was given through holes in the sapphire. Similarly, the chamber used for the phosphorylation and Hammerhead experiments was built by replacing the resin with a 254 µm thick Teflon foil from which the structure was cut out using a cutting plotter. For the phosphorylation, additional Teflon foils were placed on the warm and cold side of the chamber to mimic a hydrophobic surface. The temperature gradient was produced by heating the sapphire block through copper fixtures using heater cartridges for 3D printers and cooling the silicon side with a water bath. Temperature sensors and PID software maintained the temperatures. The temperature inside the chamber was calculated from the chamber geometry and known material constants with finite element methods (Comsol).

Fluorescence was measured with a fluorescence microscope (Zeiss Axio) through the transparent sapphire heating block using Mitutoyo infinity corrected long working distance objectives (2x, 10x) and a Zeiss Fluar 5x objective. The accumulation of DNA was detected with 200 nM FAM-labeled 132 base ssDNA in 0.1-fold PBS (see Supplementary Material for sequence). The silicon substrate was maintained at 10 °C, while the copper heaters were initially turned off. The experiment began with the copper heaters set to maintain a desired temperature. Background intensity levels were obtained from the non-fluorescing gas region. The bulk fluorescence signal was obtained from an area far from the free interface. The accumulated DNA fluorescence ratio was averaged perpendicular to the interface and divided by the bulk fluorescence. The flow was visualized with 200 nm FAM-labeled polystyrene beads in 0.1x PBS. The position and velocities of the beads were tracked using ImageJ.

**Simulation Protocol.** Simulations were performed using the finite element software COMSOL v 4.4. The 2D model solved the convective heat equation, molecule diffusion equation, and Navier-Stokes equations perpendicular to the contact line. Marangoni flows were established by implementing a stress boundary condition of the fluid velocity at the interface and by introducing a temperature-dependent surface tension. The water vapor concentration was simulated in the gas region above the interface by a diffusion-convection equation. The gas velocity was calculated from the temperature profile. Its velocity, combined with diffusion, caused an efficient net mass transport of vapor away from the interface into the gas bubble. By coupling a temperature dependent vapor concentration boundary condition to the interface, a velocity boundary condition for water at the interface was imposed by the state equation of water. This resulted in capillary flow and evaporative mass transport of the water vapor, allowing vapor to enter and escape the water through evaporation or condensation. Thermophoresis was introduced *via* a thermophoretic drift term in the convection-diffusion equation describing the DNA concentration, the Soret coefficients of the DNA were taken from experimental data[49]. To incorporate the time lag of the heating process, the sapphire temperature was measured over time and incorporated as a polynomial function for $T_{warm}$.

The above simulation was solved over time, resulting in the time evolution of accumulation at the contact line. To fit the simulation to the observed data, the surface tension dependency on temperature was set within observed values[50] and fine-tuned as a free parameter using the bead tracking data. The geometry of the interface itself was adjusted to moderately tune the DNA accumulation dynamics. Concentrations were determined by averaging the top ~10 µm of the meniscus at the hot side and comparing this to the simulated and experimental fluorescence, establishing a relation between fluorescence and average tip-concentration.

To simulate the bubble shown in Supplementary Fig. 9, the simulation was transferred to an axial-symmetric geometry with spherical coordinates. Here, the simulation around a small bubble was closed and gravity was pointed downwards. This removed the slow convection of water in the simulation, which, as expected from our modeling, did not change the accumulation characteristics. Ion and salt diffusion coefficients were taken from [51-53].

Additional experimental details can be found in the Supplementary Information.

## Data availability statement

The authors declare that the data supporting the findings of this study are available within the paper and its supplementary information files. Additional information and files are available from the corresponding author upon reasonable request. X-ray crystallographic data was also deposited at the Cambridge Crystallographic Data Centre (CCDC) under the following CCDC deposition number: 1847429. Correspondence and requests for materials should be addressed to D. B.

## Code availability

The complete details of both simulations are documented in the .html report and .mph simulation files in the Supplementary Information.

## Acknowledgments

We thank Lorenz Keil for help with data analysis. Financial support from the Simons Foundation (318881 to MWP, 327125 to DB), the German Research Foundation (DFG) through CRC/SFB 235 Project P07 and through SFB 1032 Project A04, DFG Grant BR2152/3-1, and the U.S.-German Fulbright Program is gratefully acknowledged. HM is supported by the MaxSynBio consortium, which is jointly funded by the Federal Ministry of Education and Research of Germany and the Max Planck Society. HM and KLV are supported by the Volkswagen Initiative "'Life? – A Fresh Scientific Approach to the Basic Principles of Life". AK was supported by a DFG fellowship through the Graduate School of Quantitative Biosciences Munich (QBM).

## Author Contributions

M.M., J.L., C.F.D., A.K., A.I., and Ph.S. performed the experiments, M.M., J.L., K.L.V., S.I., B.S., D.B.D., H.M., P.S., M.W.P., C.B.M., and D.B. conceived and designed the experiments, M.M., J.L., K.L.V., S.I., M.K.C., H.M., M.W.P., and D.B. analyzed the data, M.M., J.L., and D.B. wrote the paper. All authors discussed the results and commented on the manuscript.

## Competing Interests

The authors declare no competing interests.